\documentclass[a4paper,11pt]{article}

\usepackage{contribution}

\newcommand{\weblink}[2][]{%
    \ifthenelse{\equal{#1}{}}%
    {\textnormal{\url{#2}}}%
    {\textnormal{\href{#2}{#1}}}%
}

\newcommand{\acknowledgements}[1]{%
  \bigskip\bigskip
  \textsf{\textbf{\Large Acknowledgements}} \\[2ex]
  {#1}
  \bigskip
}

\def\beq{\begin{equation}}
\def\eeq#1{\label{#1}\end{equation}}
\def\eeqn{\end{equation}}

\def\beqa{\begin{eqnarray}}
\def\eeqa#1{\label{#1}\end{eqnarray}}
\def\eeqan{\end{eqnarray}}

\let\bar=\overbar

\def\Dslash{\not{\hbox{\kern-4pt $D$}}}
\def\dslash{\not{\hbox{\kern-2pt $\del$}}}

\def\msb{{\bar{\ssstyle M \kern -1pt S}}}

\newcommand{\contribution}[7][]{%
  \clearpage
  \thispagestyle{plain}
  \ifthenelse{\equal{#1}{}}
  {\hypersetup{pdftitle={#2}}}
  {\hypersetup{pdftitle={#1}}}
  \hypersetup{pdfauthor={{#3} {#4}}}
  {\centering\normalfont\LARGE\bfseries\sffamily #2 \par\nobreak}
  \lhead{}
  \chead{%
    \textit{\footnotesize XIV International Conference on Hadron Spectroscopy
      (\weblink[\textit{hadron2011}]{http://www.hadron2011.de}), 13-17 June 2011, Munich, Germany}%
  }
  \rhead{}
  \bigskip
  \begin{center}
    {#3} {#4}\ifthenelse{\equal{#6}{}}{}{\footnote{\weblink[#6]{mailto:#6}}}
    \ifthenelse{\equal{#7}{}}{}{#7} \\
    \textit{#5}
  \end{center}
  \bigskip
}

\renewcommand{\abstract}[1]{%
  \begin{center}
    \begin{minipage}{0.85\textwidth}
      \begin{footnotesize}
        #1
      \end{footnotesize}
    \end{minipage}
  \end{center}
  \bigskip
}

\begin{document}

%
%
%
%
%
{  


%

\contribution[Three charms]  
{Baryon bound states of three hadrons \\ with charm and hidden charm}  
{Chu-Wen}{Xiao}  
{$^a$ Departamento de F\'{\i}sica Te\'orica and IFIC, Centro Mixto Universidad de Valencia-CSIC,\\
Institutos de Investigaci\'on de Paterna, Aptdo. 22085, 46071 Valencia,Spain\\
$^b$ Department of Physics, Kocaeli University, 41380 Izmit, Turkey}  
{chuwen.xiao@ific.uv.es}  
{$^a$, Melahat Bayar $^{a,b}$ and Eulogio Oset $^a$}  
%

\abstract{%
  In this talk, we show our recent theoretical results for three-body systems in the charm sector which are made of three hadrons and contain one nucleon, one $D$ meson and in addition another meson, $\bar{D}$, $K$ or $\bar{K}$.
}
%

\section{Introduction}

    While the three baryon system has been a subject of intense theoretical study, it has only been recently that attention was brought to systems with two mesons and one baryon. The low lying excited $J^P=1/2^+$ $\Lambda$ and $\Sigma$ states were described in \cite{alberone}, and $N^*$ states in \cite{alberdos}, combining Faddeev equations and chiral dynamics. A $N^*$ state around 1920 MeV was predicted in \cite{jidoenyo} as a molecule of $NK \bar{K}$, corroborated in \cite{nstarpheno} and \cite{MartinezTorres:2010zv} by Faddeev equations. For three mesons systems, the X(2175) (now $\phi(2170)$) was explained as a resonant $K \bar{K} \phi$ system in \cite{albermeson}. Similarly the  $K(1460)$ is explained as a $K K\bar{K}$ state in \cite{alberjido}. 

    In a recent work we study the three body systems in the charm sector, and use the Fixed Center Approximation to the Faddeev equations (FCA), which has been proved to be reliable in \cite{Gal:2006cw,Bayar:2011qj} and has been applied to the study of the $NK \bar{K} $ system \cite{Xie2010} and the results compare favorably with those of the Faddeev approach in \cite{nstarpheno} and those of the variational approach in \cite{jidoenyo}. There are some well known two body states in this sector, such as $\Lambda_c(2595)$ in $DN$ with its coupled channels interaction \cite{lutzcharm, mizuangels}, $D_{s0}^*(2317)$ in $KD$ interaction \cite{Hofmann:2003je,Guo:2006fu,danielfirst}, and the hypothetical $X(3700)$ generated in isospin I=0 $D \bar{D}$ interaction \cite{danielfirst}. These states are the clusters in the FCA in our study.

\section{Formalism}

   Following \cite{Bayar:2011qj,Xie2010,Roca2010}, we will apply the FCA to study the charm sector. The FCA approximation to Faddeev equations is depicted in Figure~\ref{fig:fca}.
   
\begin{figure}[htb]
  \begin{center}
    \includegraphics[width=0.5\textwidth]{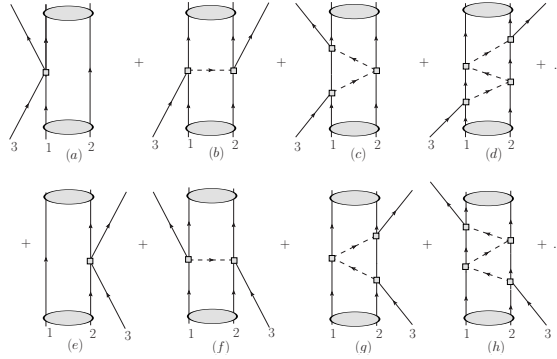}
    \caption{Diagrammatic representation of the FCA to Faddeev equations.}
    \label{fig:fca}
  \end{center}
\end{figure}
%

With this meaning of the FCA, the equations can be written by two partition functions $T_1$, $T_2$ which sum all diagrams of the series of Fig. \ref{fig:fca},
\begin{align}
T_1&=t_1+t_1G_0T_2,\label{threet1}\\
T_2&=t_2+t_2G_0T_1,\label{threet2}\\
T&=T_1+T_2,\label{threet}
\end{align}
where $T$ is the total three-body scattering amplitude. The amplitudes $t_1$ and $t_2$ represent the unitary scattering amplitudes respectively. And $G_0$ is the propagator of particle 3.

\section{Results}

   In this part we show the results of our investigation in that systems $\bar{K}DN$, $NDK$ and $ND\bar{D}$. In Figure~\ref{fig:kbdn} (left) we show the results of $|T|^2$ for the $\bar{K}\Lambda_c(2595)$ scattering in the $\bar{K}DN$ system. We find a peak around 3150 MeV, slightly above the threshold of the $\Lambda_c(2595)+\bar{K}$ mass (3088 MeV) and below the threshold of the $\bar{K}DN$ system (3298 MeV), of which the width is about 50 MeV. For the system $\bar{K}DN$, its quantum numbers are $C=+1,S=-1$ and $J^P=\frac{1}{2}^+$ since we only consider the interaction among the components in $L=0$.
\begin{figure}[htb]
  \begin{center}
    \includegraphics[width=0.49\textwidth]{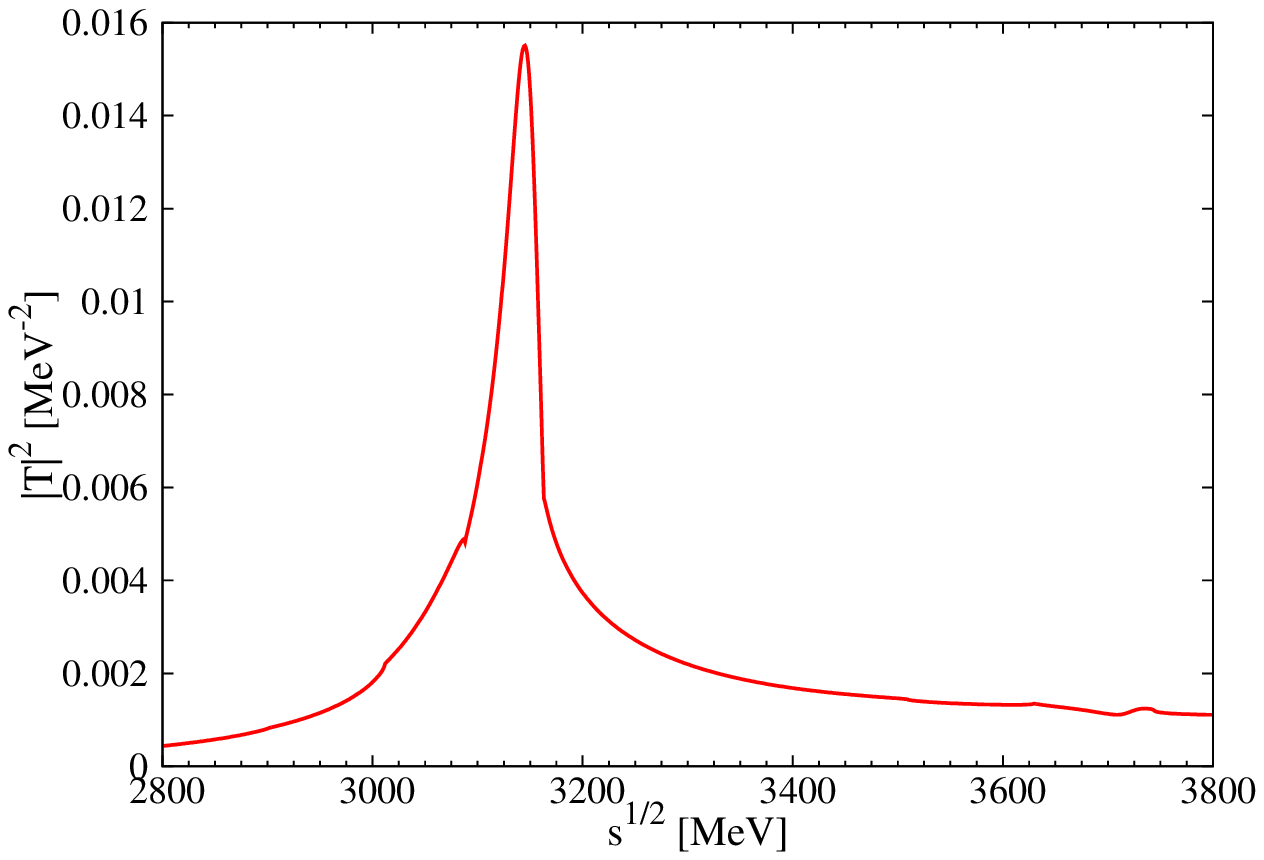}
    \includegraphics[width=0.49\textwidth]{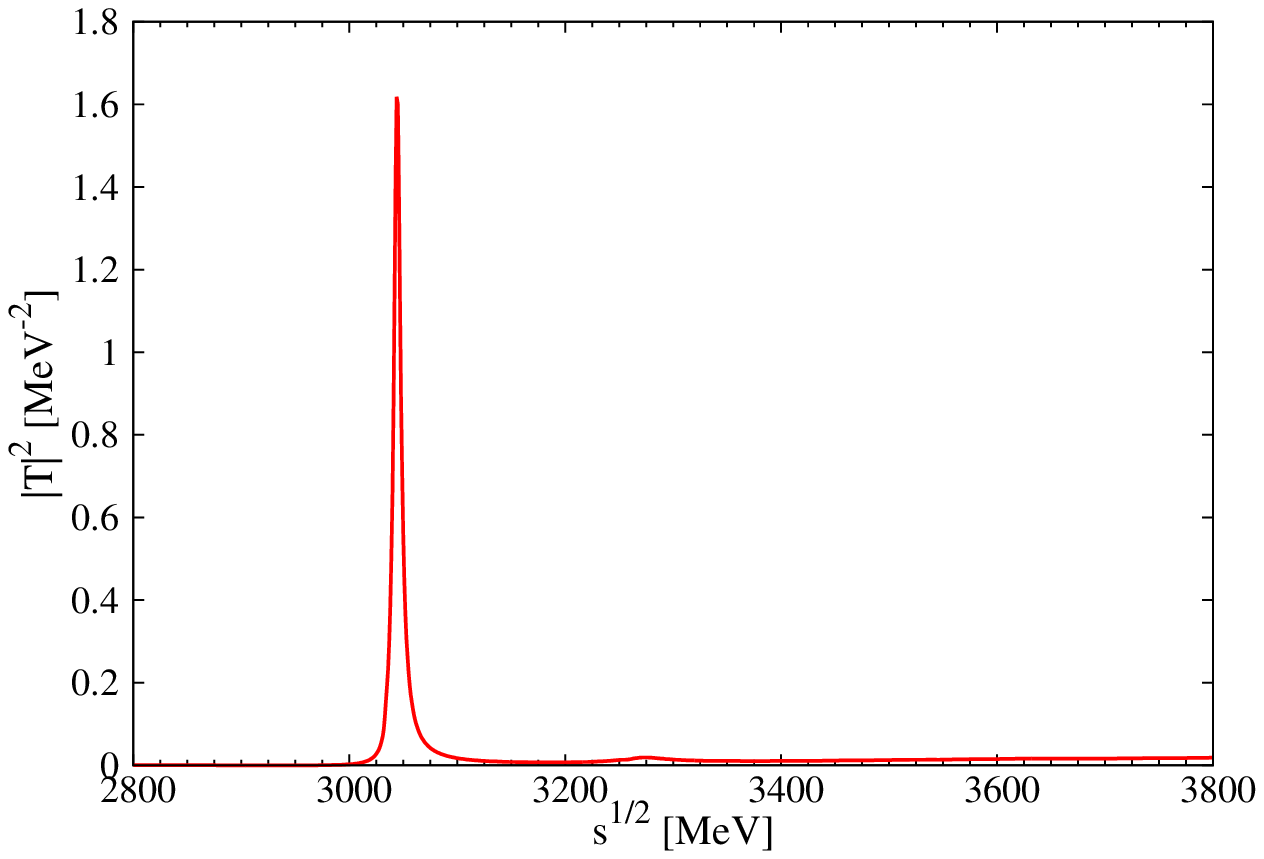}
    \caption{Modulus squared of the scattering amplitude for $\bar{K}\Lambda_c(2595)$ (left) and $ND_{s0}^*(2317)$ (right).}
    \label{fig:kbdn}
  \end{center}
\end{figure}

In the $NDK$ system, we obtain $|T|^2$ for the  $ND_{s0}^*(2317)$ scattering shown 
in Fig. \ref{fig:kbdn} (right). We found a peak around 3050 MeV which is about 200 MeV below the $N+ D_{s0}^*(2317)$ threshold and the width less than 10 MeV. We also do not 
find a counterpart in the PDG and the quantum numbers, with positive strangeness, correspond to an exotic state.
%

Finally we obtain the $T$ matrix, for the $ND\bar{D}$ interaction by means Eq. \eqref{threet}, and show the results of $|T|^2$ in Figure~\ref{fig:nddb}. From this figure we can see that there is a clear peak of $|T|^2$ around 4400 MeV and the width is very small, less than 10 MeV. The peak appears below the $ND \bar{D}$ and $N X(3700)$ thresholds and corresponds to a bound state of $N X(3700)$. This would be a hidden charm baryon state of $J^P=\frac{1}{2}^+$ which appears in the same region of energies as other hidden charm states of $J^P=\frac{1}{2}^-$ obtained in \cite{wu,wudos}.
\begin{figure}[htb]
  \begin{center}
    \includegraphics[width=0.5\textwidth]{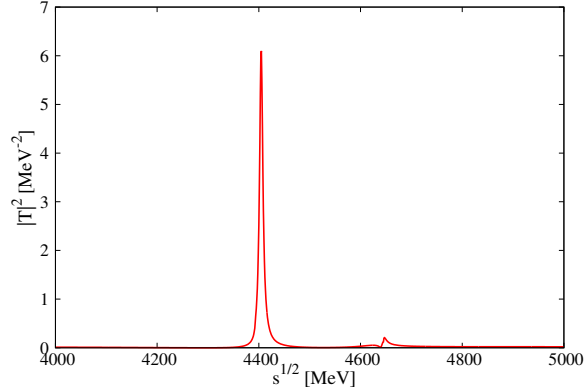}
    \caption{Modulus squared of the the $N X(3700)$ scattering amplitude.}
    \label{fig:nddb}
  \end{center}
\end{figure}

\section{Conclusion}

In all cases we find bound or quasibound states, relatively narrow, with energies 3150 MeV, 3050 MeV and 4400 MeV, respectively. All these states have $J^P=1/2^+$ and isospin $I=1/2$ and differ by their charm or strangeness content, $S=-1, C=1$, $S=1, C=1$, $S=0, C=0$, respectively. We hope that the work stimulates other theory calculations and future experiments in Facilities of FAIR or BELLE upgrade to prove our findings.

\acknowledgements{%
This work is partly supported by DGICYT contract  FIS2006-03438,
the Generalitat Valenciana in the program Prometeo and 
the EU Integrated Infrastructure Initiative Hadron Physics
Project  under Grant Agreement n.227431.
M. Bayar acknowledges support of the Scientific and Technical Research Council (TUBITAK)
BIDEP-2219 grant.
}


%

}  


\end{document}